# Simulation of Thin-TFETs Using Transition Metal Dichalcogenides: Effect of Material Parameters, Gate Dielectric on Electrostatic Device Performance

Kanak Datta and Quazi D. M. Khosru
Department of Electrical and Electronic Engineering
Bangladesh University of Engineering and Technology
Dhaka-1205
Email: kanakeee08@gmail.com

# Introduction:

Layered van der Waals materials, few layers thick or exfoliated down to single layer, have become subject of extensive research in recent times due to their exciting electronic and optoelectronic properties [1, 2]. Apart from Graphene, Transition Metal Dichalcogenides (TMDC), with tunable bandgap ranging from 1eV to 2 eV, have shown great potential as channel materials for next generation transistor applications [3][4][5][6][7]. With the ever growing concern for power dissipation in large scale CMOS circuits for next generation high speed microprocessors, Tunneling Field Effect Transistors (TFET) are considered as attractive and viable candidates due to their ability to beat the fundamental 60 mV/dec subthreshold limit of conventional MOSFETs. TFETs invoke the physics of interband tunneling which allows them to achieve very steep subthreshold characteristics and therefore low off-state leakage [8]. TFETs based on III-V semiconductors, which are attractive materials due to lower carrier effective mass, direct energy bandgap, ease of heterostructure and heretojunction formation with advanced growth techniques, have been studied and investigated for quite some time [9], [10]. More recent investigation on TFETs involves the inclusion of 2D semiconductors which are considered promising candidates for ultra-scaled devices due to their scalability down to monolayer level thickness, improved interface properties compared to existing bulk materials. With scalability down to sub-nanometer level, devices based on TMDC monolayers or stacked TMDC multilayer heterojunctions can achieve excellent electrostatic control and sharp turn 'on' characteristics. Moreover, lack of dangling bonds at the interface allows the growth of stacked TMDC device structures with low interface state defects which can be exploited in high performance TFET application [11].

Implementation of scaled tunneling field effect transistors based on 2D materials requires type-II (staggered gap) or type-III (broken gap) band alignment that effectively increases electron tunneling efficiency at 'on' state and suppresses leakage at 'off' state of the device [12]. This requirement of broken gap band alignment can be effectively fulfilled by using different combination of 2D materials. Atomistic simulation on TMDC materials has revealed wide tunability in their bandgap and electron affinity values which can be exploited to design high performance tunneling devices for next generation low power microprocessors [13], [14]. Electronic properties of these materials can be further modulated by applying strain and vertical electric field [15], [16]. Apart from monolayer 2D materials, different TMDC monolayers stacked on top of each other i.e. bilayer heterostructure based on 2D materials allow further tunability in electronic and optoelectronic properties [17], [18]. Moreover, calculations using Density Functional Theory (DFT) have shown that mixed ternary 2D compounds may be synthesized with thermodynamic stability at room temperature [19]. The DFT calculation also shows that, the direct bandgap nature of the TMDC monolayers is maintained in their alloy compounds, which makes these materials suitable for optoelectronic applications [19]. Besides theoretical investigation, lateral and vertical heterostructures based on TMDC monolayer alloys have also been reported in [20].

In this work, we have performed a numerical study on thin n-TFETs [11], [21] based on 1T-SnSe$_2$. We have formed different bilayer TFET structures using monolayer 1T-SnSe$_2$ and other 2H-MX$_2$ (M=Mo/W; X=Se/Te) monolayer TMDC materials and performed numerical simulation using the transport model explained and reported in [11][21]. As mentioned earlier, for the successful realization of TFETs, it is required to have type-II or type-III band alignment at the heterojunction interface. Both these types of heterojunction can be obtained at different TMDC material interfaces. Here, in our particular device, we have used the top TMDC layer as the n-type material layer. Therefore for operation as n-TFET, we need the top layer to have a large electron affinity. Looking at affinity values of TMDC materials reported in recent literature, we can see that, 1T-SnSe$_2$ has a high value of electron affinity [21]. Therefore, when paired with other TMDC monolayers like WSe$_2$, MoTe$_2$ etc. SnSe$_2$ monolayer can obtain the required type-II or type- III band alignment at the interfaces for successful n-TFET operation.

In our calculation, we have used material parameters reported in recent literature. At first, we have performed 1D self-consistent simulation assuming long channel approximation to determine the conduction band and valance band energy levels at different 2D layers in the device as a function of gate bias voltage. After that, the interlayer tunneling current is determined using a simplified transport model reported and explained in [11]. Later on, performance parameters like 'on' current, subthreshold-swing, threshold voltage are explored. Moreover, we have also performed device simulations using TMDC binary alloys as the bottom layer material and compared their performances.

## **Device Structure:**

The device structure used in this work is proposed by Li et al. [11], which is shown in fig. 1. The proposed device has two layers of monolayer TMDC materials stacked on top of each other separated by an interlayer material. The top and bottom monolayers are connected to drain and source terminals respectively. The doping at top and bottom layer has been considered to be n-type and p-type respectively. The electrostatics of the device is controlled by two gates (top and bottom) as shown in fig 1. The bottom gate is considered to be grounded ($V_{BG}=0$) throughout this numerical study. We apply voltage only at the top gate ($V_{TG}$). In our work, we have used same material as top and bottom gate oxides for general analysis. The tunneling of electrons take place from the bottom layer to the top layer and the device is turned on when the conduction band minima of the top layer falls below the valance band maxima of the bottom monolayer. The tunneling mechanism is shown in fig. 2 in a simple manner. In addition, we have also investigated the device performance when the bottom layer is made of monolayer TMDC alloys, MXY (M=Mo/W; X=S/Se; Y=Te).

## Simulation Process:

In this work, we have divided the simulation process into two parts: self-consistent determination of the conduction and valance band positions in the device and calculation of the interlayer tunneling current. We have employed the long channel approximation for determining conduction and valance band level using self-consistent analysis. We have solved Poisson's equation in 1-D (along gate confinement direction i.e. x- direction) as shown in Fig. 1. The tunneling current has been calculated for 1-D case only. Analysis of two-dimensional tunneling that may extend beyond the ends of the respective layers is beyond the scope of this work. A more accurate model of transport characteristics should include the two dimensional effects that becomes very important in ultra-scaled devices. Consideration of two dimensional effects would degrade device electrostatics, increase leakage current and therefore lead to degraded sub-threshold device performance. For transport calculation, we have employed one band model i.e. considered a single band transport for band to band tunneling. The same formulation has also been employed in [11].

- **Determination of Band Profile Using Self-Consistent Method:**

Using the values of bandgap, electron affinity, effective mass of TMDC monolayers reported in recent literature, we have solved the 1-D Poisson equation self-consistently to determine the conduction and valance band profile inside the device. For simplified analysis, we have

employed the long channel approximation in our calculation. Poisson's equation in 1-D can be written as:

$$\nabla^2 \psi(x) = -\frac{\rho(x)}{\varepsilon}; \quad E_C(x) = -q\psi(x) \quad and \quad E_V(x) = E_C(x) - E_g \tag{1}$$

Where $\psi(x)$ refers to the potential profile and $\rho(x)$ refers to the charge density inside the device at different regions. To calculate the carrier density in the device we have used the semi-classical formulation as used in [11]:

$$n(x) = \frac{g_c m_c k_B T}{\pi \hbar^2} \ln[\exp(-\frac{q\phi_{n,T}(x)}{k_B T}) + 1]; \quad \phi_{n,T}(x) = E_{CT}(x) - E_{FT} \tag{2}$$

$$p(x) = \frac{g_v m_v k_B T}{\pi \hbar^2} \ln[\exp(-\frac{q\phi_{p,B}(x)}{k_B T}) + 1]; \quad \phi_{p,B}(x) = E_{FB} - E_{VB}(x) \tag{3}$$

$$\rho(x) = e(N_d + p(x) - N_a - n(x)) \tag{4}$$

Here, $g_c/g_v$ refer to degeneracy factor in the conduction/valance band, $m_c/m_v$ refer to conduction/valance band effective masses of the carrier, $E_{FT}$ and $E_{FB}$ refers to the Fermi level of the top and bottom layer TMDCs respectively. The Fermi level of bottom TMDC is considered to be grounded, $n(x)$ and $p(x)$ refer to electron and hole concentration in the top and bottom TMDC layers respectively, $N_d$ and $N_a$ refer to the doping at the top and bottom TMDC layers respectively. From eqn. 4, we can determine the carrier concentration in the top and bottom 2D layers. These charge concentrations are fed into eqn. 1 and the self-consistent loop continues. After convergence is achieved, we obtain the conduction and valance band levels inside the device.

- **Transport Equation:**

In this work, for transport calculation, we have used the formalism for single electron transport developed in [11] which has been used to model vertical 2D material monolayer heterjunction transistors [22]. According to this formalism, the single particle tunneling current can be expressed as [11]:

$$I = g_v \frac{4\pi e}{\hbar} \sum_{\mathbf{k}_T, \mathbf{k}_B} |M(\mathbf{k}_T, \mathbf{k}_B)|^2 \delta(E_B(\mathbf{k}_B) - E_T(\mathbf{k}_T))(f_B - f_T) \quad [5]$$

Where, $e$ is the electronic charge, $\mathbf{k}_B$ and $\mathbf{k}_T$ are the wave vectors of the bottom and top 2D material layers respectively, $E_T(\mathbf{k}_T)$ and $E_B(\mathbf{k}_B)$ are corresponding energies at top and bottom layers, $f_B$ and $f_T$ are Fermi occupation function at the top and bottom layers, $g_v$ is the valley degeneracy. $M(\mathbf{k}_T, \mathbf{k}_B)$ refers to the matrix element that expresses the charge transfer between the two layers of 2D material. According to eqn. 5 the tunneling current depends on the scattering matrix element. Therefore, proper calculation of the matrix element is necessary to accurately extract the tunneling current between the top and bottom TMDC layers which would require the exact nature of wave function in the top and bottom layers and the nature of the potential variation in the middle interlayer region [11]. Therefore this matrix element can be different for different material combination used in the TFET.

However, this above mentioned matrix element can be simplified using semi-classical calculation under the assumption that, the conduction band minima and valance band maxima occur at the same point in the Brillouin zone and the difference between wave vectors at the two layers i.e. $\mathbf{q} = \mathbf{k}_T - \mathbf{k}_B$ is very small compared to the size of the Brillouin zone [11]. As shown in [11] we can write the matrix element as:

$$|M(\mathbf{k}_T,\mathbf{k}_B)|^2 \approx \frac{|M_{B0}|^2 S_F(\mathbf{q})}{A} e^{-2\kappa T_{IL}} \qquad [6]$$

Where, $\kappa$ is the decaying constant of the wavefunction in the interlayer, $T_{IL}$ refers to the thickness of the interlayer. $S_F(\mathbf{q})$ is the power spectrum of the scattering potential which relaxes the momentum conservation and allows tunneling for $\mathbf{k}_B \neq \mathbf{k}_T$ [11][21]. The power spectrum of the scattering potential can be written as [11]:

$$S_F(q) = \frac{\pi L_C^2}{(1+q^2 L_C^2/2)^{3/2}} \qquad [7]$$

Where, $L_C$ refers to the correlation length, which has been set at 10 nm in [11][21] and q= |**q**|. In our work, we have set $L_C$ at 10 nm as well. With the above mentioned considerations, the transport equation can be written as:

$$I = g_v \frac{e|M_{B0}|^2 A}{4\pi^3 \hbar} e^{-2\kappa T_{IL}} \int_{\mathbf{k}_B} \int_{\mathbf{k}_T} d\mathbf{k}_T d\mathbf{k}_B S_F(\mathbf{q}) \delta(E_B(\mathbf{k}_B) - E_T(\mathbf{k}_T))(f_B - f_T) \qquad [8]$$

Here, the squared matrix element $|M_{B0}|^2$ can be related to interlayer charge transfer time across the van der Waals gap between the two TMDC layers [11]. According to eqn. 6, the tunneling current is proportional to the matrix element $|M_{B0}|^2$ and decreases exponentially with interlayer thickness and decay constant. Eqn. 8 can be further modified to account for the finite energy broadening phenomenon which induces a tail in the density of states characteristics and affect the abrupt turn 'on' characteristics of the tunneling device. To account for the finite energy broadening effect, in [11][21], in place of $\delta(E_B(\mathbf{k}_B)-E_T(\mathbf{k}_T))$, a energy broadening spectrum, $S_E(E_B(\mathbf{k}_B)-E_T(\mathbf{k}_T))$, is used:

$$S_E(E_T(\mathbf{k}_T) - E_B(\mathbf{k}_B)) = \frac{1}{\sqrt{\pi\sigma^2}} e^{-(E_T(\mathbf{k}_T) - E_B(\mathbf{k}_B))/\sigma^2}$$

[9]

Here, the energy broadening spectrum has been approximated using a Gaussian function with $\sigma$ being the energy broadening parameter in the 2D material layers [11][21]. Finally the tunneling current equation becomes:

$$I = g_v \frac{e|M_{B0}|^2 A}{4\pi^3 \hbar} e^{-2\kappa T_{IL}} \int_{\mathbf{k}_B} \int_{\mathbf{k}_T} d\mathbf{k}_T d\mathbf{k}_B S_F(\mathbf{q}) S_E(E_B(\mathbf{k}_B) - E_T(\mathbf{k}_T))(f_B - f_T)$$

[10]

In this work, we have assumed the value of $M_{B0}$ to be 0.01 eV. As mentioned before, the correlation length, $L_C$ is assumed to be 10 nm. The energy broadening in the top and bottom layer has been assumed to be 10 meV. In real semiconductors, presence of impurities, dopant distribution, phonon assisted transport, lattice imperfection can result in energy level broadening which results in an exponential tail like edge in the absorption spectrum. This tail is known as the Urbach tail [23]. This tail like nature in real semiconductors is presented by Urbach parameter $E_0$. The Urbach parameter is a strong function of temperature and increases with increasing temperature [23, 24]. For semiconducting systems like a-Si, the value of Urbach parameter has been found to be close to 40~50 meV . However, for GaAs, due to weak electron-phonon interaction, the Urbach parameter has been found to be unusually very small and the bandtail i.e. the absorption tail shows a simple exponential dependence with $E_0(T)$ increasing almost linearly with T [24, 25]. At room temperature (300K), the Urbach parameter for GaAs in semi-insulating and Si-doped state have been found to be close to 7 and 12.5 meV respectively [24]. In reference to Li et al. [11], we have used the energy broadening value of 10 meV in our simulation. We have simulated our device structures with different values of energy broadening and found similar results reported in [11].

In this work, the interlayer thickness between top and bottom TMDC has been assumed to be 0.6 nm. The dielectric constant in the intermediate layer is considered to be 1.0. Furthermore, the tunneling current also depends on the decay constant of the wavefunction in the middle layer. In [21], the value of the decay constant, $\kappa$ has been calculated assuming a barrier height, $E_B$ of 1 eV and free electron mass, $m_0$ in the interlayer. Using $\kappa = \sqrt{\frac{2E_B m_0}{\hbar^2}}$, the value of decay constant was found to be 5.12 nm$^{-1}$. We have used the same value of decay constant in our calculation. However, changing the dielectric material in the interlayer region would result in a change in the bandgap and therefore the effective barrier height at the TMDC- interlayer dielectric interface which would eventually affect the decay constant in the interlayer. A more accurate model of carrier transport should therefore include this change in bandgap while evaluating decay constant and calculating the interlayer band to band tunneling. In this work, for transport calculation of thin TFETs, the spin-orbit coupling effect in the valance band has not been taken into account. The valance band has been considered to be spin-degenerate. In addition, for simplified transport analysis, the possibility of rotational misalignment between the two TMDC layers has not been taken into account in this work. The effect of rotational misalignment on the transport characteristics of thin TFETs has been carried out in [11]. Rotational misalignment between top and bottom TMDC layers would induce a rotational matrix in the calculation of matrix element M($k_B$,$k_T$) and would lead to a broadening of the scattering spectrum. However, analytical calculations in [11] suggest that, such rotational misalignment may affect the absolute value of tunneling current without affecting the gate voltage dependence of the tunneling phenomenon. The effects of disorder, lattice imperfections and other interactions like electron-phonon interactions have been taken into account by the energy broadening function presented in eqn. 9. Enhanced electron-phonon coupling would lead to increased energy broadening and therefore degraded subthreshold performance of the device. However, the 'on' current of the device as found in this work and in Li et al.[11] should not be affected by increased electron-phonon coupling.

In ultra-scaled short channel devices, source-drain access geometry can significantly affect the transport characteristics of the device. Moreover, depending on the TMDC material used, access geometry, material disorder and other imperfections, the channel access resistance would be different and therefore the lead-limited or source-drain lead limited current would be different for same bias conditions. In this work, we have simulated our device in long channel regime. Therefore, source-drain resistance effect may be considered negligible for our device. Moreover, as shown in Li et al. [1], the inclusion of source-drain access resistance may only affect the drain current at high gate bias regime and the sub-threshold behavior of the device may not be affected. The drain current would decrease with increasing resistance at high gate bias voltage.

## **Material Parameters Used:**

The material parameters used in this study along with their references are given in TABLE 1 [21][26][27][28][29][30]. For the study using TMDC alloys as the bottom layer materials we have used the parameters given in TABLE 2. In ref [31], bowing parameters are given for bandgap, conduction and valance band positions. For other parameters like electron and hole effective mass, we have calculated the parameters using simple Vegard's law and the calculated parameters are given in TABLE 3.

The values of effective mass and dielectric constant of $WTe_2$ monolayer has been extracted from [32]. In our analysis, we have assumed n-type and p-type doping density of $10^{12}$ /cm$^2$ in top and bottom TMDC layers respectively. In all our simulations, we have considered the top gate and bottom gate metal work-functions to be 5.6 eV and 4.8 eV respectively. Similar values of gate metal work-functions were also used in [21] for the simulation of n-TFETs.

## Results and Discussion:

In this section, we will discuss the results obtained from our simulation of thin TFETs based on monolayer TMDC materials. We will also discuss the results obtained using monolayer TMDC alloys as bottom layer material. In all these simulations, unless mentioned otherwise, we have kept the bottom gate voltage, $V_{BG}$ kept fixed at 0V.

- **Monolayer TMDC TFETs:**

As mentioned earlier, for monolayer case, the top TMDC is 1T-SnSe$_2$ and the bottom TMDC layer can be MoSe$_2$, WSe$_2$ or MoTe$_2$. In our basic simulation, we have considered SiO$_2$ as the top and bottom gate dielectric material. The dielectric constant of the interlayer material has been considered as 1.0.

Fig. 3a shows the valance band and conduction band positions as a function of the top gate voltage for SnSe$_2$/WSe$_2$ TFET. As seen from the figure, the top gate clearly exerts control over the top TMDC conduction band level. The effect of top gate on the valance band energy of the bottom TMDC layer can be seen from the figure as well. As top gate voltage increases, the valance band of the bottom layer rises above the conduction band of the top TMDC layer and the tunneling window opens up constituting current flow in the device. Fig. 3b shows the electron density in the top TMDC layer as a function of top gate bias voltage. As seen from the figure, with increasing top gate voltage, the electron density in the top TMDC layer increases which marks the electrostatic effect of the top gate resulting in carrier accumulation in device operation under long channel approximation.

Fig. 4a shows the drain current as a function of drain bias voltage at different top gate bias voltage conditions. Thin-TFETs show good electrostatic effect of the top gate as can be seen from the figure. The drain current gets saturated and remains almost constant with increased drain bias voltage. Fig. 4b shows the drain current as a function of energy broadening in the top and bottom TMDC layers in the device. In our basic simulation, we have considered the energy broadening to be 10 meV in both top and bottom layers. From fig. 4(b) with an energy broadening of 10 meV we have extracted a Subthreshold Swing (SS) of 17 mV/dec. The figure also shows the Id-$V_{TG}$ characteristics as we change the energy broadening parameter in our simulation. As discussed before, in real devices, the energy broadening caused by electron-impurity and electron-electron interaction in a many body environment, can limit the abrupt turn 'on' characteristics of the device and therefore can affect the subthreshold swing. This effect has been discussed in greater detail in [11]. We have found similar results in our simulation as well. As we increase the energy broadening term, we have observed linear increase in the subthreshold device current and degraded SS values. The increase in SS with energy broadening is shown in the inset figure of Fig. 4b. This linear trend of SS increment with energy broadening is also reported in [11].

Fig. 5 shows the Id-$V_{TG}$ characteristics of the thin TFET for three different bottom TMDC materials- $MoSe_2$, $WSe_2$, $MoTe_2$ in log scale. The energy broadening parameter has been considered to be 10 meV in this case. As can be seen from the figure, for similar top and bottom layer doping density and gate metal work function and interlayer thickness, $MoSe_2$ can provide higher threshold voltage among the bottom layer materials. On the other hand, MoTe$_2$ appears to be providing the lowest threshold voltage. From the reported values of conduction band level with reference to vacuum level i.e. electron affinity in [13], we find that WSe$_2$ has the lowest affinity values of the three materials while MoSe$_2$ has the highest. From the values of electron affinity and bandgap, it is clear that, MoSe$_2$ and WSe$_2$ both would form type-II band alignment. However, due to higher bandgap of WSe$_2$, SnSe$_2$/ WSe$_2$ TFET would have a lower threshold voltage than SnSe$_2$/MoSe$_2$ TFET. On the other hand, MoTe$_2$, with lower affinity value and bandgap would form type-III band alignment with the valance band of bottom MoTe$_2$ positioned above the conduction band of top SnSe$_2$ monolayer. Therefore, we see the lowest threshold voltage for MoTe$_2$ monolayer TFET. The devices show very little difference in subthreshold characteristics as can be seen from the figure.

We have also simulated the device structure with variable interlayer thickness and variable dielectric constant in the interlayer. In these studies, we have used SiO$_2$ as the top and bottom gate dielectric material. We have used 1T-SnSe$_2$ as the top layer TMDC material and 2H-WSe$_2$ as the bottom layer TMDC material. Fig. 6a shows the conduction and valance band positions in the device as a function of the top gate voltage $V_{TG}$ for different values of dielectric constant of the middle layer. As seen from the figure, the cross-over point between valance and conduction band positions shifts toward higher top gate voltage as we increase the value of dielectric constant. This ultimately leads to an increase in the threshold voltage of the device. With increment in dielectric constant, the interlayer capacitance increases, which makes the energy levels at the bottom TMDC layer more sensitive to applied voltage at the top gate of the device. This adversely affects the subthreshold device performance and degrades the sharp turn 'on' characteristics of the device. Fig. 6b shows the Id-$V_{TG}$ characteristics of the device as a function of the top gate voltage, $V_{TG}$ when the interlayer dielectric constant is varied. As can be seen from the figure, with increasing dielectric constant, the subthreshold behavior of the device could degrade significantly and subthreshold swing could increase with increasing interlayer dielectric constant. The 'on' current however is not affected much by the change in interlayer dielectric constant. The figure also shows an increase in threshold voltage with increment in interlayer dielectric constant.

Fig. 7a shows the conduction and valance band positions at different top gate voltage values when interlayer thickness is varied. Changing interlayer thickness causes similar effect seen as before (Fig. 6a) as it changes the interlayer capacitance. With lowered interlayer thickness, the interlayer capacitance increases and the cross-over point shifts towards higher top gate voltage value. Fig. 7b shows the Id-$V_{TG}$ characteristics when interlayer thickness is varied. From eqn. 10, we see that the tunneling current has an inverse exponential relationship with the interlayer thickness. Therefore, when the device is fully turned 'on', lower interlayer thickness would lead to higher drain current. On the other hand, lowering interlayer thickness increases interlayer capacitance and increases subthreshold swing of the device.

We have also explored the effect of variable top gate dielectric on device performance. In basic simulation so far, we used $SiO_2$ as both the top and bottom gate dielectric material. In the study of variable top gate dielectric, we have used $SiO_2$ as the bottom gate dielectric material. Fig. 8a shows the effect of top gate dielectric on the energy levels in the device. As seen from the figure, the cross-over point between valance and conduction band appears roughly at the same point of the applied top gate voltage. Fig. 8b shows the Id-$V_{TG}$ characteristics of the device at different top gate oxide conditions. As can be seen from the figure, the threshold voltage is not affected by the change in top gate dielectric material. However, the top gate dielectric does affect the subthreshold device performance just like in normal MOSFETs. High-k dielectric material gives better subthreshold performance and lowers subthreshold swing. However, the 'on' current in this case appears to be not affected.

- **Effect of Top Gate Workfunction:**

In our basic simulation, we have considered the top gate metla work function to be 5.6 eV and bottom gate metal workfunction to be 4.8 eV. However, we have also simulated the effect of top gate metal workfunction on device performance. The effect of top gate metal workfunction variation is shown in fig. 9. As can be seen from the fig. 9(a), with increasing top gate metal workfunction, the crossover point between the conduction band of the top layer and the valance band of the bottom layer gradually shifts towards higher and higher voltage which increases the device threshold voltage. The increment of threshold voltage with top gate metal workfunction increment can also be observed from fig. 9(b). However, the subthreshold characteristics of the device remains unaffected with top gate metal workfunction variation.

- **Monolayer TMDC Alloy TFETs:**

We have also simulated TFETs with monolayer TMDC alloys used as the bottom layer material. Using parameters reported in [31], we have followed the same simulation procedure and have extracted the conduction and valance band energy levels in the device and determined drain current. In our simulation, we have explored four such material systems-MoSTe, MoSeTe, WSTe and WSeTe.

Fig. 10 shows the Id-$V_{TG}$ characteristics of the simulated device at different bottom layer TMDC alloys for similar top and bottom gate metal work-functions. As can be seen from our simulation, like monolayer TMDC TFETs, TFETs with TMDC alloys at the bottom layer show similar trend in threshold voltage. Of the systems used in this simulation, SnSe$_2$/WSeTe TFET shows the lowest threshold voltage and SnSe$_2$/MoSTe TFET shows the highest threshold voltage.

## Conclusion:

In this work, we have performed a numerical simulation study on thin, n-TFET using different TMDC monolayer materials. For simplified transport analysis, a single band transport model reported is recent literature is used along with long channel approximation. Rotational misalignment between the two TMDC layers is beyond the scope of this work. The energy broadening phenomenon, that is critical to the sharp turn-on behavior of the TFET, has been approximated using a Gaussian function whose validation can be found in literature. This Gaussain energy function incorporates effects of disorders, lattice imperfections, defects, and other non-idealities that contribute to the subthreshold performance of the TFET. The effect of interlayer dielectric has been explored considering decay of electronic wavefunction in the interlayer material. A more accurate modeling of transport characteristics would require exact nature of electronic wavefunction in the top and bottom TMDC layers and also in the interlayer regime of the device. As shown from our simulation, the threshold voltage of the device depends strongly on the electron affinity and bandgap of the material system being used in the TFET. An increase in device subthreshold swing was observed with increased energy broadening which has also been reported in recent literature. The implementation of high-k gate dielectric material could improve the subthreshold characteristics of the TFET without significant effect on the 'on' current of the device. The interlayer material also plays a vital role in the device performance, as seen from this numerical study. Lowering interlayer thickness improves charge transfer by increasing interlayer tunneling and therefore leads to higher device current. However, it makes the bottom layer energy levels more sensitive to top gate voltage and affects subthreshold device performance adversely. Interlayer dielectric constant also affects device performance. Material with higher dielectric constant increases interlayer capacitance which causes increment in device threshold voltage. However, like reduced interlayer thickness, it also increases subthreshold swing. The simulation on TMDC alloys also reveal similar dependence of device performance on electron affinity and material bandgap. Therefore, optimization and proper combination of material structures, gate dielectric and interlayer materials are necessary in designing steep subthreshold TMDC TFETs that would meet the future requirement of next generation low power microprocessors.